\documentclass[rnote,referee]{aa}
\usepackage{times,graphicx}

\begin{document}


\title{Tensorial depolarization of   alkali atoms by isotropic collisions with neutral hydrogen.
}



\author{M. Derouich}

\institute{
}
\titlerunning{Collisional relaxation}
\authorrunning{M. Derouich}

\date{Received 01 June 2012 / Accepted 23 July 2012}
\abstract
{
We consider the problem of isotropic collisions between an alkali atom and neutral hydrogen. We  calculate the collisional tensorial components of general p and s-states, characterized by their effective principal quantum number $n^{*}$. It is found that the behaviour of the tensorial components obey simple  power  laws allowing quick calculations of the depolarizing collisional rates. As application, our results should allow  a rigorous treatment of the atomic polarization profiles of the  D1 -D2  lines of alkali atoms.
 }
{Close coupling treatments of atomic collisions are needed to decipher the information encoded in the polarized radiation from the Sun. Important problems remain unresolved like the role of collisions in the Paschen-Back conditions.}
 \\

\maketitle

\section{Context, notations and numerical results}
\subsection{Context}
Atomic polarization consists in an unbalance of the populations of the Zeeman sublevels pertaining 
to a given atomic level and in the presence of coherences between the sublevels themselves. 
The scattering polarization is the observational signature of the atomic polarization (e.g. Landi Degl'Innocenti \& Landolfi 2004; Trujillo Bueno 2001). In real plasmas like the solar atmosphere, emitting atoms suffer the effects of wide variety of collisions with nearby abundant particles. Sometimes the information encoded in accurate observations would be inaccessible if the effect of the collisions is misunderstood or ignored. 

Scattering polarization is sensitive to radiation, 
to magnetic and electric fields, to collisions, etc. As the scattering polarization is usually rather 
small in the problems of practical interest (e.g. scattering polarization of the order of 1\% in Fraunhofer lines 
in the photosphere and chromosphere of the Sun), careful and precise modelling of that polarization is of fundamental importance for learning especially about a weak and even unresolved 
magnetic field by its Hanle effect. For this purpose, it is of interest to obtain detailed statistical equilibrium equations (SEE) 
including all the tensorial components of the collisional rates. 
\subsection{Notations}
Let us  consider  the atomic states ($\alpha$ $j$)    of  alkali atoms where   $j$ is the total angular momentum   and  $\alpha$ represents 
 the other quantum numbers necessary to  define the electronic level.    The atomic states ($\alpha$ $j$) are quantified  by the spherical tensor components $\rho_{q}^{k} (\alpha j)$  of the density matrix, $k$ is the tensorial order and $q$ represents the coherences between the levels (e.g.\ Fano 1963, Omont 1977,  Sahal-Br\'echot 1977; Landi Degl'Innocenti  \& Landolfi 2004). 
It is useful to note that for isotropic processes the coupling terms implying $k \ne k'$ and transfer of coherence $q$ to $q'$ are zero, and that the collisional depolarization and polarization transfer rates are $q$-independent. At the solar photosphere where the second solar spectrum is formed and at the low chromosphere,   the dominant collisions with neutral hydrogen are isotropic,  so that the non-diagonal components of the collisional depolarization matrix are zero.   
 
 In the case of collisions of neutral atoms with neutral hydrogen, the inelastic and super-elastic excitation between two different electronic levels  are negligible.  Therefore, the indice $\alpha$  is omitted  from now on for the sake of brevity. We apply the general theory developed by Derouich et al. (2003, 2005) and the associated numerical code to provide, the all non zero tensorial collisional rates of any $s$ and $p$ level.  We denote by $C^{k}_E$   the collisional rates  due to elastic collisions  and by $D^{k}(j)=C^{0}_E(j)  - C^{k}_E(j)$ the usual depolarizing rate; the expression of $D^{k}$ is given for example by equations (7) and (9) of Derouich et al. (2003).  $C^k_I(j,j_l)$  and  $C^k_S(j,j_u)$ are the collisional transfer rates due to inelastic and super-elastic collisions respectively (see Derouich et al. 2007; Landi Degl'Innocenti \& Landolfi 2004).  
The indices $l$ and $u$ denote any level of energy  respectively lower or higher than the energy of the level ($j$).
The contribution of isotropic collisions in the SEE is given by Equation (5) of Derouich et al. (2007) and Equation (7.101) of Landi Degl'Innocenti \& Landolfi (2004).

\subsection{Numerical results}
 The present work provides new complementary numerical results indispensable for rigorous analysis of the polarization profiles.     Derouich et al. (2003) determined solely the depolarization rates $D^{k=2}$ of P-states. In fact, Table 4   of Derouich et al. (2003) gives the alignment ($k=2$) depolarization rates as a function of $n^*$ for the temperatures of 5000 K and 6000 K. The  collisional rates $C^{k}_E$, $C^{k}_I$ and $C^{k}_S$  were not given.  This paper  attempts to fill this lacuna by providing  $C^{k}_E$, $C^{k}_I$ and $C^{k}_S$  as a power function of $n^*$ and $T$.  In addition,   compact analytical power laws of the $C^{k}_E$ rates associated to S-states are   derived   from the collisional method presented in Derouich et al. (2005).

Grids of collisional rates are computed for  the effective principal quantum 
number $n^*$ of the  s and p-states in the interval [1.5, 3].  
 We adopt a step size of 0.1 and we obtain    laws behaviours with $n^*$ which are, with correlation coefficients $R >0.9$, fit by: 
\begin{description}
\item[---]
\textbf{Spherical tensor components of the level $3s$ $^2S_{j=\frac{1}{2}}$}
\begin{eqnarray}  \label{eq_11}
C^{0}_E(\frac{1}{2})  &=&  1.838 \times 10^{-9} \;  n_{\textrm {\scriptsize H}} \times   n^{*^{3.675}} \left(\frac{T}{5000}\right)^{0.416} \nonumber \\
C^{1}_E(\frac{1}{2})  &=&   0.958 \times 10^{-9} \;  n_{\textrm {\scriptsize H}} \times   n^{*^{4.011}}   \left(\frac{T}{5000}\right)^{0.416}   
\end{eqnarray}
\item[---]  \textbf{Spherical tensor components of the level $3p$ $^2P_{j=\frac{1}{2}}$}
\begin{eqnarray}  \label{eq_11}
C^{0}_E(\frac{1}{2})  &=&  1.301 \times 10^{-9} \;  n_{\textrm {\scriptsize H}} \times   n^{*^{2.293}} \left(\frac{T}{5000}\right)^{0.406} \nonumber \\
C^{1}_E(\frac{1}{2})  &=&   0.753 \times 10^{-9} \;  n_{\textrm {\scriptsize H}} \times   n^{*^{2.279}}   \left(\frac{T}{5000}\right)^{0.420}  
\end{eqnarray}
\item[---]  \textbf{Spherical tensor components of the level $3p$ $^2P_{j=\frac{3}{2}}$}
\begin{eqnarray}  \label{eq_11}
C^{0}_E(\frac{3}{2})  &=&  1.685 \times 10^{-9} \;  n_{\textrm {\scriptsize H}} \times   n^{*^{2.671}} \left(\frac{T}{5000}\right)^{0.420} \nonumber \\
C^{1}_E(\frac{3}{2})  &=&  0.953 \times 10^{-9} \;  n_{\textrm {\scriptsize H}} \times   n^{*^{2.773}} \left(\frac{T}{5000}\right)^{0.390} \nonumber \\
C^{2}_E(\frac{3}{2})  &=&  0.472 \times 10^{-9} \;  n_{\textrm {\scriptsize H}} \times   n^{*^{2.400}} \left(\frac{T}{5000}\right)^{0.380} \nonumber \\
C^{3}_E(\frac{3}{2})  &=&  0.482 \times 10^{-9} \;  n_{\textrm {\scriptsize H}} \times   n^{*^{2.670}} \left(\frac{T}{5000}\right)^{0.400}   \\
C^{0}_I(\frac{3}{2}, \frac{1}{2})  &=&  0.975 \times 10^{-9} \;  n_{\textrm {\scriptsize H}} \times   n^{*^{2.705}} \left(\frac{T}{5000}\right)^{0.400} \nonumber \\
C^{1}_I(\frac{3}{2}, \frac{1}{2})  &=& - 0.095 \times 10^{-9} \;  n_{\textrm {\scriptsize H}} \times   n^{*^{2.687}} \left(\frac{T}{5000}\right)^{0.410} \nonumber
\end{eqnarray}
Only the excitation  collisional transfer rates $C^{0}_I(\frac{3}{2}, \frac{1}{2})$ and $C^{1}_I(\frac{3}{2}, \frac{1}{2}) $ are given. However, it is 
straightforward to retrieve the values of the   deexcitation  collisional rates 
 $C^{0}_S(\frac{1}{2}, \frac{3}{2}) $  and  $C^{1}_S(\frac{1}{2}, \frac{3}{2}) $ by applying the detailed balance relation:
 \begin{eqnarray}  \label{eq_11}
C^k_S(j,j_u) &=&     \frac{2j+1}{2j_u+1}  \exp \left(\frac{E_{j_u}-E_{j}}{k_BT}\right) \; \; C^k_I(j_u,j)
\end{eqnarray}
with $E_{j}$ being the energy of the level ($j $) and $k_B$ the Boltzmann 
constant. 
\end{description}
\begin{table} 
\begin{center}
\begin{tabular}{l c c c c c c c c c c c r}
\hline
  &   Li I & Na I & K I &  Rb I &   Cs I   \\
n$^*$ ($^2$S) & 1.588 &1.627 &1.770 &1.804 &1.869\\
n$^*$ ($^2$P) & 1.960 & 2.117 & 2.234 & 2.285 & 2.329 \\
\hline
\end{tabular}
\end{center}
\caption{Effective quantum number n$^*$ of the alkali atoms.}
\label{V-nondiagonal}
\end{table}

We notice that all the collisional rates are given in s$^{-1}$, $n_{\textrm {\scriptsize H}}$ is the neutral hydrogen density in cm$^{-3}$ and the temperature $T$ is in Kelvins. 
Table 1 gives the effective principal quantum number $n^*$ of the s and p-states of the alkali atoms Li I, Na I, K I, Rb I and Cs I. In particular, since   $n^*$=1.627 for ground state of Na I,  the destruction of orientation rate ($k=1$) is  
\begin{eqnarray}  \label{eq_11}
D^1(\frac{1}{2}) &=&  C^{0}_E(\frac{1}{2}) -C^{1}_E(\frac{1}{2})  \nonumber \\
&=& \left[1.838 \times   n^{*^{3.675}} - 0.958 \times   n^{*^{4.011}}\right] \nonumber \\ && \hspace{1cm} \times 10^{-9} \;  n_{\textrm {\scriptsize H}}   \left(\frac{T}{5000}\right)^{0.416}   \\
  &=&   4.246  \times 10^{-9} \;  n_{\textrm {\scriptsize H}} \times    \left(\frac{T}{5000}\right)^{0.416}    \nonumber 
\end{eqnarray}
which is very close to the fully quantum rate given in the erratum of Kerkeni et al. (2000)  (At 
$T$=5000K, they found a rate of 4.32  $\times$ 10$^{-9}$ $n_{\textrm {\scriptsize H}}$ s$^{-1}$).
\section{Hyperfine structure}
In typical solar conditions, the hyperfine splitting is usually much lower than the inverse of 
the typical time duration of a collision and therefore one can assume that the nuclear spin is 
conserved during the collision\footnote{It is important, however, not to confuse this condition with the fact that the  SEE
 must be solved for the hyperfine levels when the inverse of the lifetime of the level is lower than the 
hyperfine splitting, i.e. the hyperfine levels are separated.}.  This is the frozen nuclear spin approximation implying  that the 
hyperfine collisional rates are given as a linear combination of the fine rates   $C^{k}_E$, $C^{k}_I$ and $C^{k}_S$ given in this paper
 (e.g. Nienhuis 1976 and Omont 1977). 

\section{Conclusion}
 Tensorial colisional components are given in this work and could be implemented in the numerical simulations of 
the scattering polarization in a way similar to radiative rates. Hyperfine and fine collisional rates 
may be derived from simple power laws provided in this work. 

Our results are valid for collisions without external magnetic field or for a sufÞciently weak 
field. In fact, under solar conditions, to the best of the author's knowledge, there have been no study which includes  magnetic fields explicitly in the 
calculations of collisional rates. Depolarizing  collisional rates  commonly encountred in the literature  should  be used in the Hanle effect regime 
but are not valid in the 
Paschen-Back effect regime.  It remains a challenge to take into account the effect of the magnetic fields in the calculation of the 
 collisional cofficients .

\end{document}